\documentclass{vldb}
\usepackage{graphicx}
\usepackage{subfigure}
\usepackage{balance}  
\usepackage{amsmath}
\usepackage{amssymb}
\usepackage{algorithmic}
\usepackage{algorithm}
\usepackage{pifont}

\begin{document}


\title{Efficient Destination Prediction Based on Route Choices with Transition Matrix Optimization}



%
%
%
%

\author{
%
%
Zhou Yang\textsuperscript{\dag}\titlenote{Corresponding author.}, Heli Sun\textsuperscript{\dag}, Jianbin Huang\textsuperscript{\ddag}, Xiaolin Jia\textsuperscript{\dag}\\ Ziyu Guan\textsuperscript{\S}, Zhongmeng Zhao\textsuperscript{\dag}\\
\textit{\textsuperscript{\dag}Department of Computer Science and Technology, Xi'an Jiaotong University, Xi'an, China} \\
\textit{\textsuperscript{\ddag}School of Software, Xidian University, Xi'an, China} \\
\textit{\textsuperscript{\S}School of Information and Technology, Northwest University, Xi'an, China} 
}

\maketitle

\begin{abstract}
Destination prediction is an essential task in a variety of mobile applications. In this paper, we optimize the matrix operation and adapt a semi-lazy framework to improve the prediction accuracy and efficiency of a state-of-the-art approach. To this end, we employ efficient dynamic-programming by devising several data constructs including Efficient Transition Probability and Transition Probabilities with Detours  that are capable of pinpointing the minimum amount of computation. We prove that our method achieves one order of cut in both time and space complexity. The experimental results on real-world and synthetic datasets have shown that our solution consistently outperforms its state-of-the-art counterparts in terms of both efficiency (approximately over 100 times faster) and accuracy (above 30 \% increase).
\end{abstract}

\section{Introduction}
Location prediction is a central theme of mobile computing. It has found its wide applications in domains such as providing smooth handoffs between wireless communication cells, offering cognitive assistance, providing additional information for in-car navigation system and so on \cite{krumm2006predestination, patterson2004opportunity, xue2013destination, horvitz2012some, evensen2011mobile}. The problem can be defined as: given the current location $c$ and the starting location $s$ of a partial trip already traveled by a user, we seek to find out the destination $d$ of the whole journey.

\begin{figure}
	\centering
	\includegraphics[scale=0.72]{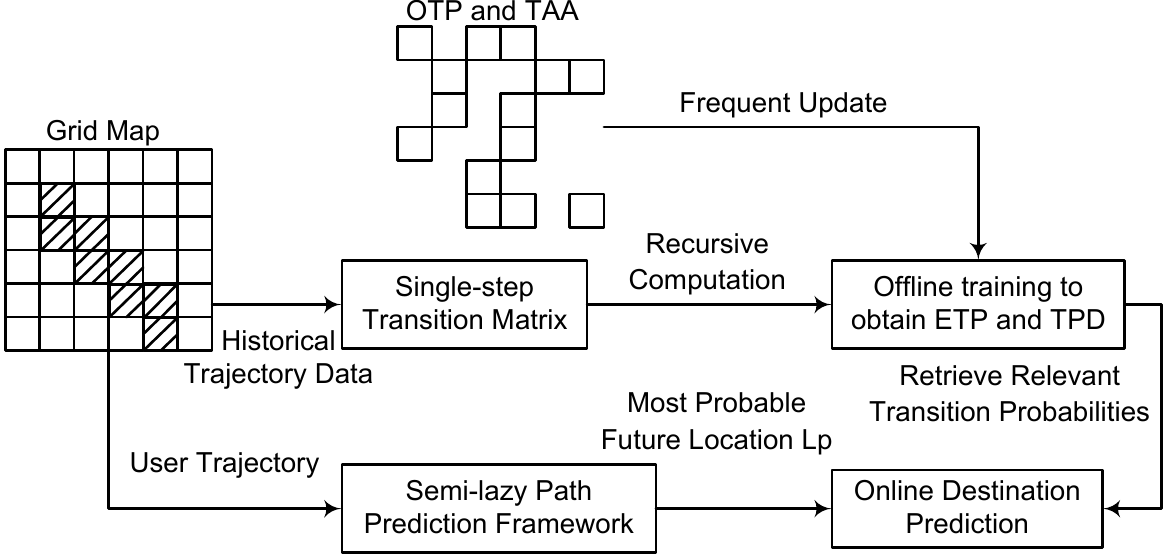}
	\caption{The components of our approach.}
	\label{OverviewSketch}
\end{figure}

Our destination prediction scheme consists of the following components (Figure \ref{OverviewSketch}). First, two novel data constructs Efficient Transition Probability (ETP) and Transition Probability with Detours (TPD) are employed to efficiently train the offline prediction model. ETP and TPD can pinpoint the minimum indispensable computation. Then we locate Obligatory Transit Point (OTP) and Transition Affected Area (TAA) to efficiently update the preceding offline model (only recompute the altered transition probabilities). OTP and TAA further impose constraints on the minimum region of interest (hotspot). Subsequently we use a semi-lazy method to identify the most probable future location regarding the recent route choice of a user. This strategy is applied in conjunction with the Bayesian theory-based online prediction to improve the prediction results.

Our research is motivated by the fact that
 most earlier work\cite{krumm2006real,krumm2006predestination,xue2015solving,xue2013desteller,ziebart2008navigate}
 
(1) cannot deal with the data spasity problem very well. Some locations are never covered in the historical data. The lack of data reduces the prediction accuracy of these methods.

(2) underutilize the available historical data. Most previous work builds a model with the data and very often this model is not a lossless representation of the original one.

(3) rather notable efficiency improvement of algorithm can be gained through our optimization. We note that matrix multiplication involved in our baseline can be simplified through our dynamic-programming like approach. Moreover, frequent update of the model can be achieved by restraining the computation to the minimum amount combined with this mechanism.

The main contributions of our work are summarized as below.

(1) We incorporate the semi-lazy framework into our prediction model to address the problem overlooked by most earlier work  \cite{krumm2006real,krumm2006predestination,xue2015solving,xue2013desteller,ziebart2008navigate} to adequately consider the route choice between the starting location and the current location.

(2) We propose an efficient dynamic-programming like algorithm which includes two flexible data constructs -- ETP and TPD -- that vastly improves the training efficiency.

(3) We devise effective mechanisms OTP and TAA to deal with the alteration of transition probabilities confined merely to several regions (around 5\% of the total amount). We then exploit the efficiency gain for more frequent update of our model to improve prediction accuracy.

The rest of the paper is organized as follows. We introduce the related work in this domain and give an overview of our approach in section 2. After that, the adaptation of semi-lazy prediction, the optimization of matrix multiplication, harnessing the effect of detour distances, efficient frequent update of the prediction model, are laid out in section 3, section 4 and section 5 respectively. In section 6, we present the experiments. Finally, we conclude our work in section 7.

\section{Related Work}
Previous work pertaining to this subject tends to discover patterns which they term as “popular” for subsequent decision-making that seeks to optimize a certain goal, be it best routes connecting two endpoints 
\cite{gonzalez2007adaptive,luo2013finding,su2014crowdplanner,wei2012constructing,chen2011discovering,zheng2014modeling}, the next most probable stops or the region that appears to be of interest to drivers
\cite{do2012contextual,jeung2008hybrid,mathew2012predicting,monreale2009wherenext,yuan2012discovering}.

Krumm and Horvitz \cite{krumm2006predestination} incorporated in their approach multiple features including driving efficiency, ground cover, trip times etc. and employed an open-world model to capture the probabilities of users leaving for places which have never been visited in the past. Ziebart et al. \cite{ziebart2008navigate} employed a sophisticated context-aware behavior model PROCAB to infer intersection decisions, routes and destinations of drivers. Gao et al. \cite{gao2014elastic} demonstrated the breach of privacy induced by insurance schemes through predicting the destinations by exploiting only the speed data of a vehicle. The aim of our work differs from all the preceding research since we concentrate on destination prediction using only the historical trajectories.

Jeung et al. \cite{jeung2008hybrid} proposed two query processing techniques that can obtain future movement prediction through a novel access index. The knowledge of the possible end points of a journey also facilitates opportunistic routing, conceived by Eric Horvitz et al. \cite{horvitz2012some}, that recommends sensible diversions along one trip route to a primary destination. Monreale et al. \cite{monreale2009wherenext} designed a T-pattern Tree which is learnt from Trajectory Patterns. Future trajectory prediction \cite{sametKDD16aircraft} has even been applied in the Decision Support System (DST) of Air Traffic Management (ATM). Trinh Minh Tri Do et al. \cite{do2012contextual} developed an ensemble method that builds a contextual variable into a probabilistic framework for human mobility prediction. These studies are directed at the prediction of subsequent movement, next places or the future trajectory all of which do not address the problem of destination prediction as our work does.

A majority of the aforementioned research focuses on one or several geo locations that matter most, either the current position or some statistically significant places, and then perceive them as several discrete states of a Markov model or a HMM \cite{do2012contextual,horvitz2012some,krumm2006predestination,mathew2012predicting,xue2015solving,xue2013destination}.

Our approach prioritizes the most recent movements. This is an apparent advantage over our baselines \cite{xue2015solving} which essentially consider only the starting location and the current location. In fact, Xue et al. \cite{xue2013destination} utilizes this trait for privacy protection against their SubSyn algorithm by removing two endpoints (i.e. the origin and the current position).

Our work relies solely on the historical trajectory dataset to predict destinations, which is notably different from most previous work \cite{do2012contextual,horvitz2012some,krumm2006predestination,su2014crowdplanner,ziebart2008navigate} (i.e. no other information such as time or user profiles is included). This general setting allows us to analyze user movement when such knowledge is not available, which is often the case. 

\begin{table}
\scriptsize
	\centering
	\caption{The adopted notations.}
	\begin{tabular}{|c|p{5.5cm}|} \hline
		\textbf{Symbol} & \textbf{Detail}\\ \hline
		$\mathrm{\theta }$  & The confidence threshold strikes the balance between the length of $\mathrm{D}_p$ and prediction  accuracy \\ \hline
		$\mathrm{D}$  & Total trip distance \\ \hline 
		$d_t$ & Distance traveled so far \\ \hline 
		$\mathrm{D}_p$ & The length of predicted path \\ \hline 
		$\mathrm{\alpha }$ & Decay factor \\ \hline 
		$d_i$ & Subinterval boundary point (distance)  \\ \hline 
		$\mathrm{T}_p$ & The trajectory traveled so far  \\ \hline
		$\mathrm{L}_p$ & Most probable future location \\ \hline 
		$d_p$ & The length of predicted path of the semi-lazy prediction framework \\ \hline 
		$\mathrm{P}_d$  & Destination probability \\ \hline 
		$p_{a\to b}$ & Total transition probability for a trip from location $a$ to location $b$  \\ \hline 
		$\mathrm{P(d|s)}$ & Transition probability for a trip starting at location $s$ and ending at location $d$  \\ \hline 
		$M^l$ & $l$-step Markov transition matrix  \\ \hline 
		$l_{ab}$ & The L1 distance between location $a$ and location $b$  \\ \hline 
		$\mathrm{T}_s$ & Trajectories starting at $s$ \\ \hline 
		$\mathrm{T}_{s,d}$ & Trajectories starting at $s$ and ending at $d$ \\ \hline 
		$l_d$ & Length of detour \\ \hline
	\end{tabular}
	\label{table1Notations}
\end{table}

\section{Adaptation of THE SEMI-LAZY PREDICTION FRAMEWORK}
In this section, we discuss about the adaptation and incorporation of the semi-lazy framework into our prediction model during the online training phase. We first determine the predicted length of ongoing trajectory and then identify the location that is most likely to be traveled in the future. After that our model produces the predicted results given the knowledge of this most probable future location. The notations used in this paper are given in Table \ref{table1Notations}.

\subsection{The Workflow}
\begin{figure}[tbph]
	\centering
	\includegraphics[scale=1.5]{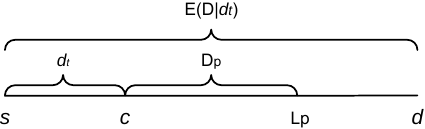}
	\caption{The relationship between all the concerning variables.}
	\label{PredictionLine}
\end{figure}

The basic workflow (Figure \ref{PredictionLine}) of our adaptation is as follows. First we employ the semi-lazy path prediction framework \cite{zhou2013semi} to generate a path that connects the current location $c$ to the most probable future location $\mathrm{L}_p$. We specify the desirable predicted length $\mathrm{D}_p$ through applying a logarithmic decay to $\mathrm{E}\left(\mathrm{D}|d_t\right)$, the estimated total trip distance at the current timestamp. Then the end point of $\mathrm{D}_p$ (i.e. $\mathrm{L}_p$) replaces the current location $c$ since $\mathrm{L}_p$ is more likely to be closer to the final destination and thus gives better prediction results.

\subsection{The Predicted Path}

\textbf{Trip Estimation.} To decide on a proper value for the length of a predicted path, we first create a frequency diagram depicting the distribution of total trip distance of our historical data. Specifically, we let distance measurements fall into one of the subintervals separated by $d_i$. The expected value of total travel distance is calculated by

\begin{equation}
	\mathrm{E}\left(\mathrm{D}\right)\mathrm{=}\sum_i{d_iP(d_i<\mathrm{D<}d_{i+1})}.
\end{equation}

Then we iteratively estimate the total trip distance at a particular instant of time as

\begin{equation}
\setlength{\abovedisplayskip}{0pt}
\setlength{\belowdisplayskip}{0pt}
	\mathrm{E}\left(\mathrm{D}|d_t\right)=\sum_{i}^{\infty}{d_i\frac{P(d_i<\mathrm{D}<d_{i+1})}{P(\mathrm{D}\geqslant d_t)}},
\end{equation}

where $d_i$ should satisfy
\[
\begin{cases}
d_i<d_t,& \ i=\mathrm{sup}\mathrm{}\{i\geqslant 0|d_t>d_i\}\\
d_i\geqslant d_t,& \text{otherwise}
\end{cases}.
\]

Equation 2 provides the expected value of total trip distance given the current trip that has been traveled so far. It offers us a ballpark figure of the journey distance at a specific time which can be used to determine the predicted length $\mathrm{D}_p$. Predicted length $\mathrm{D}_p$ imposes a constraint on the upper bound of the length of trajectories generated by the semi-lazy framework which requires us to specify a proper threshold $\mathrm{\theta }$.

\textbf{Logarithmic Decay.} As mentioned in the overview of this chapter, we seek to identify $\mathrm{L}_p$ and thus only a certain proportion of $\mathrm{E}(\mathrm{D}|d_t)$ (i.e. $\mathrm{D}_p$) is taken to achieve this aim. Moreover, the rationale for the reduction of predicted path is that $\mathrm{L}_p$ should also approach the current location $c$ as the trip gradually comes to its end. Hence we employ a logarithmic function to perform this task which is given as

\begin{equation}
	\mathrm{D}_p=\mathrm{E}\left(\mathrm{D}|d_t\right){{\mathrm{log}}_{\alpha } \frac{d_t}{\mathrm{E}(\mathrm{D}|d_t)} },
\end{equation}

where the argument of the logarithm $\frac{d_t}{\mathrm{E}(\mathrm{D}|d_t)}$ quantifies the estimated trip completion percentage based on our preceding $\mathrm{E}\left(\mathrm{D}|d_t\right)$. The base $\mathrm{\alpha }$, the decay factor, indicates how fast the predicted percentage should decline. We repeatedly alter the decay factor $\mathrm{\alpha }$ in our experiment and find that setting it to 0.004 works the best.

\textbf{Translation Between Predicted Path and Confidence Threshold.} Once we have selected a proper value for $\mathrm{D}_p$, the confidence value can then be known which is in proportion to $\mathrm{D}_p$. According to \cite{zhou2013semi}, a longer path produces a lower confidence value.

The semi-lazy framework compares the confidence value and the confidence threshold $\mathrm{\theta }$ to determine the length of the predicted path. We modify the semi-lazy path prediction algorithm by incrementally comparing the length of its inferred path and that of our $\mathrm{D}_p$ to suit our need. The pseudo code of our approach is presented in Algorithm 1.

\begin{algorithm}[tbph]
                \algsetup{linenosize=\small}
                \scriptsize      
	\caption{ EDP (Efficient Destination Prediction)}         
	\label{algHEDP}
	\textbf{Input:} the trajectory traveled by a user $T_p$\\
	\textbf{Output:} the predicted destination
	
	\begin{algorithmic}[1]
		\STATE Estimate total trip distance $\mathrm{E}\left(\mathrm{D}|d_c\right)$. (Equation 2)
		\STATE Obtain $\mathrm{D}_p$ by applying logarithmic decay. (Equation 3)
		\WHILE{$d_p\mathrm{\le }\mathrm{D}_p$}
		\STATE Semi-lazy trajectory prediction generates a longer path $d_p$ with lower confidence value
		\ENDWHILE
		\STATE Identify the most probable future location $\mathrm{L}_p$
		\STATE{}\COMMENT{Destination prediction based on Markov transition probabilities.}
		\STATE $P_d\propto \frac{p_{L_p\to d}P\left(d|s\right)}{p_{s\to d}}$
	\end{algorithmic}
\end{algorithm}

$p_{L_p\to d}$ denotes the total transition probability for a journey from location $L_p$, the most probable future location, to a presumed destination $\mathrm{d}$. Likewise, $p_{s\to d}$ represents a trip which starts at $\mathrm{s}$. Note that $p_{a\to b}=M^{l_{ab}}_{ab}$ is actually an element of an $l_{ab}$-step Markov transition matrix which we obtain through multiplying the single-step matrix $l_{ab}$ times. $P\left(d|s\right)=\frac{\left|T_{s,d}\right|}{\left|T_s\right|}$ reflects the proportion of trajectories that begin at the same origin $s$ (denominator $\left|T_{s}\right|$) but end up at different locations $d$ (numerator $\left|T_{s,d}\right|$).

\section{Optimizing the Markov Transition Matrix Multiplication}

\subsection{The Motivation for Optimization}
Markov transition matrix multiplication remains to be the major hurdle of performance improvement for our offline training. In our case, matrix multiplication can be a computationally formidable task particularly when the size and the number of step of transition required by the Markov transition matrix are large. According to Xue et al. \cite{zhou2013semi}, the offline training for SubSyn takes beyond 1 hour for a map of medium grid granularity setting on a commodity machine.

\subsection{Efficient Transition Probability}

First let us introduce some key concepts.

\textbf{Definition 1} (ETP -- Efficient Transition Probability) Given two locations $i$ and $j$, $\mathrm{ETP}(i, j, l)$ is the probability of the transition taking the most efficient route whose length corresponds to the L1 distance $l$.

\textbf{Definition 2} (Relative Adjacent Pair - RAP) Given two locations $i$ and $j$, the relative adjacent pair, $A_{ij}=\{A^1_{ij},\ A^2_{ij}\}$, comprises precisely two cells that are immediately adjacent to $j$ regarding $i$ in the L1-metric sense . These two adjoining cells are on the route that links $i$ with $j$. 

A special case arises when two cells are in the same row/column. In this case the RAP comprises soley one element which is the adjoining cell of $j$ regarding $i$.

\begin{figure}
	\includegraphics[scale=0.51]{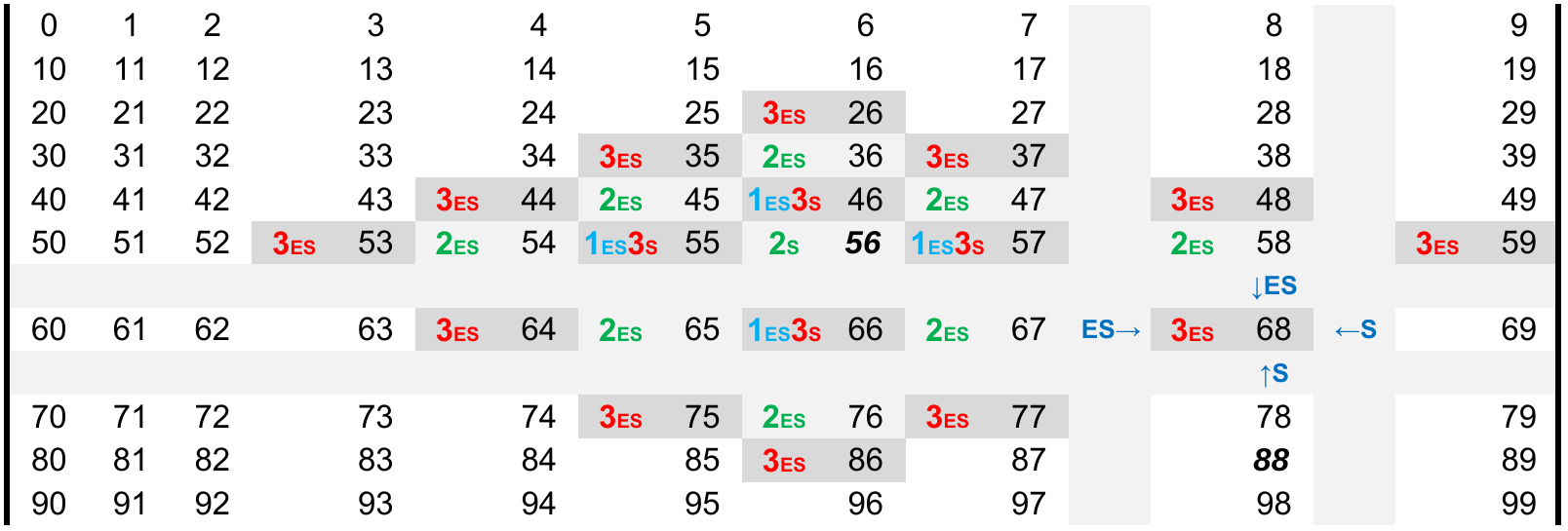}
	\caption{The elements involved in the matrix multiplication of SubSyn series algorithms (denoted by S) and our algorithm (denoted by E). The different color indicates the distinct totality of elements after one step of transition (i.e. step 1, 2 or 3).}
	\label{Figure2SubsynElements}
\end{figure}

Simply put, setting off from location $i$ one must pass either of the two cells of RAP to reach location $j$. (Say in Figure \ref{Figure2SubsynElements}, $i=56,   j=88$ then $j_1=78,   j_2=87$) This notion is essential in our solution to the cut of computational cost of one order of magnitude since it corresponds to the efficient routes taken and circumvent the extra computation brought about by sparse matrix multiplication -- two cells which are impossible to reach in this case.

\textbf{The Computation of ETP.} Next we show how to compute ETP (Efficient Transition Probability) through dynamic-programming like recursion. The relationship between $(l-1)$-step transition and $l$-step transition can be found by

\begin{equation}
\begin{split}
	\mathrm{ETP}(i,j,l)=\mathrm{ETP}(i,A^1_{ij},l-1)\times \mathrm{SSTP}(A^1_{ij},j) \\+\mathrm{ETP}(i,A^2_{ij},l-1)\times \mathrm{SSTP}(A^2_{ij},j)
\end{split}.
\end{equation}

Here $\mathrm{SSTP}\left(i, j\right)\mathrm{=}\frac{\left|T_{i,j}\right|}{\left|T_i\right|}$, the Single Step Transition Probability, measures the frequency of transition from $i$ to $j$. Equation 4 consists of exactly the two components of RAP (Relative Adjacent Pair), i.e. $A^1_{ij}$ and $A^2_{ij}$ to recursively obtain the efficient transition probabilities. The strength of this technique compared with sparse matrix multiplication is its ability to calculate the necessary transition probabilities only once and save them for later computation. It is akin to the divide-and-conquer tactic in that every problem (reaching location $j$) can be worked out by dealing with its sub problems (reaching the RAP of location $j$).

\subsection{Detour Distances}

Although L1 distance takes into consideration the case when the shortest Euclidean route is constantly infeasible, scenarios may emerge where drivers intentionally opt for a slightly longer itinerary. Despite the fact that Xue et al. \cite{xue2015solving} has excluded detour distances from their approach claiming that the exclusion leads to greater simplicity and little degradation in prediction accuracy, we report around 7\% increase in prediction accuracy with the consideration of detour distances.

\textbf{Definition 4} (TPD - Transition Probabilities with Detours) Given two locations $i$ and $\mathrm{j}$, $\mathrm{TPD}(i,j,l_d)$ gauges the probability of transition taking the route with a detour whose length is $l_d$.

The definition of TPD resembles that of ETP since $\mathrm{TPD}\left(i,j,l_d\right)\mathrm{=ETP}(i,j,l)$ when no detour is involved, i.e. $l_d=l$. Next we can find that TPD can be obtained recursively by

\begin{equation}
	\mathrm{TPD}\left(i,j,l_d\right)=\sum^4_{k=1}{\mathrm{TPD}(i,j_k,l_{ij_k})\mathrm{SSTP}(j_k,j)},
\end{equation}
where $j_k$ denotes the 4 cells that are immediately adjacent to $j$. This data construct has the capability to suit the needs of detours of different lengths without being prone to performance reduction as is matrix multiplication. Shown in Figure \ref{Figure2SubsynElements} are the locations surrounding the starting point (cell 56) alternating between two states, either reachable or unreachable.

\subsection{The Upper Bound}

It is apparent that during each step of transition one can travel to only half of the locations in the neighboring region of the starting point. This implies that at least 50\% of the transition probabilities calculated by matrix multiplication are destined to be zero. However, the techniques employed by \cite{xue2015solving,xue2013destination} require to update every element of the transition matrix during its multiplication, be it dense \cite{xue2013destination} or sparse \cite{xue2015solving}, which essentially incurs more than double the necessary cost of both computation and storage. Here by claiming ``more than double'', we are referring to the fact that at least 50\% of the entries of a transition matrix are zero which can indeed be stated as a theorem below.

\textbf{Theorem 1}. Given a $g$-by-$g$ $\mathrm{s}$-step Markov transition matrix $m$, the amount of non-zero entries is $nz=\mathrm{\{}m_{ij}\mathrm{|}m_{ij}\mathrm{\neq }\mathrm{0}\mathrm{\}}$, then $\frac{\left|nz\right|}{\left|m\right|}\le 0.5\ (\left|m\right|=g^2)\ $.

\textbf{Proof}. From any location in a map one can travel to the 4 directly adjacent cells. The ensuing move should land him on any of the 4 adjoining cells of his previous move. In order to get to his subsequent destination, he has to first leave his previous starting location, which essentially rules out his arrival on these places (starting points) in this turn of transition (self-transition excluded). Furthermore, since none of the immediately adjacent 4 cells is reachable after the previous step of transition, these starting cells cannot be got to from the other locations during this turn of transition as well (non self-transition also excluded now). The preceding reasoning applies to every step of transition, precisely rendering at least half of the total locations (starting points) impossible to get to and the other half reachable (destinations). Hence the conclusion can be drawn that the non-zero elements constitute no more than 50\% of a transition matrix. A more mathematically rigorous treatment of this issue is offered in the Appendix. $\square$

\section{Frequent update of the model}

\subsection{The Justification for Frequent Update}
 
Traffic conditions undergo instantaneous changes at every moment. It is rational that we capture its latest trend by frequently updating our model. We notice that only a portion of all the transition probabilities varies during a short period of time, for instance at three-minute intervals. To factor into such changes of road traffic, previous approaches \cite{xue2015solving,xue2013destination} have to perform matrix multiplications for all of the cells residing in a map. However, according to our analysis, since changes occur at merely a part of the whole map, redundant computations are carried out by this solution. Our approach differs from its predecessor in that it breaks down the structure of Markov transition matrix and is directed at the items that are integral to the transitions between cells in a map (ETP and TPD). Consequently, we can adjust the proportion of updates for transition probabilities to meet the requirements of the constantly changing traffic condition while keeping the incidental cost regarding these alterations as low as possible.

\subsection{Transit Points and Affected Areas}

First we present some definitions concerning the frequent update of the model.

\textbf{Definition 5} (OTP - Obligatory Transit Point) One needs to travel past the obligatory transit point $k$ on the route from location $i$ to location $j$.

\textbf{Definition 6} (TAA -- Transition Affected Area) Departing from location $i$ and walking past the intermediate transit point $j$, one is likely to reach any of the cell residing in the transition affected area $\mathrm{TAA}(i,j,\mathrm{D,T})$ after he takes his ensuing moves that may include a detour $d\mathrm{\in }\mathrm{D}$ adding up to the distance $t\mathrm{\in }\mathrm{T}$ of the whole trip. Moreover, now the intermediate transit point $j$ is exactly the OTP of any of the entire constituents of TAA.

\begin{figure}[tbph]
\centering
\includegraphics[scale=0.7]{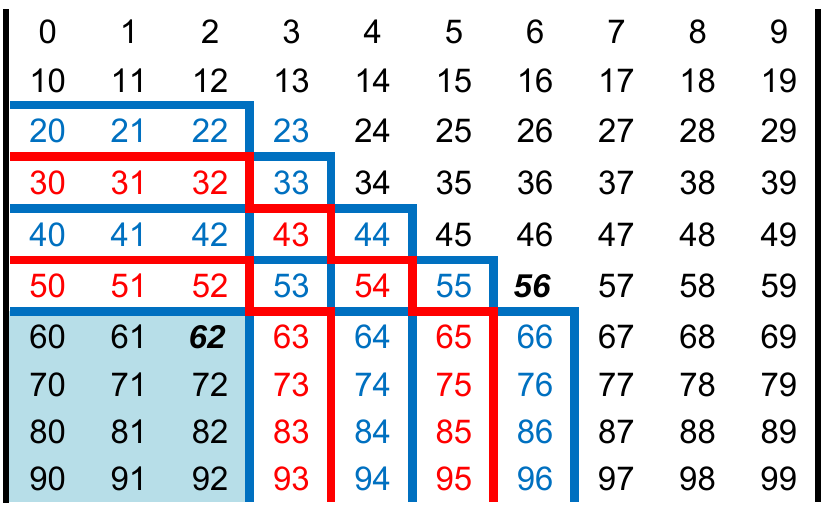}
\caption{The region of TAA( 56, 62, D, T ), D = \{0, 2, 4, 6, 8\}, a longer detour corresponds to a more extended area.}
\label{Figure7TAAandTAAExtended}
\end{figure}
Consider the following example: one starts from location 56 and makes a stop at location 62 which we also perceive it as the OTP of the $\mathrm{TAA}(56,62,\mathrm{D,T})$. For some reason the changing traffic results in the alteration of the transition probabilities of location 62 with regard to its four locations in the vicinity. Then it is evident that $\mathrm{TAA}(56,62,\mathrm{0,T})$ is composed of the cells forming the rectangle whose diagonal spans from location 62 to location 90 (Figure \ref{Figure7TAAandTAAExtended}).

$\mathrm{TAA}(56,62,\mathrm{2,T})$ expands beyond the area of $\mathrm{TAA}(56,62,\mathrm{0,T})$, further occupying the adjoining cells of its top and right border (Figure \ref{Figure7TAAandTAAExtended}). Such expansion occurs with the increase of the length of detour, enabling us to identify the TAA in a recursive fashion (i.e. to compute the values of ETP in each TAA). Notice that the increment of the third item of the tuple TAA, the detour $\mathrm{D}$, should always be 2 according to Theorem 1.

\subsection{The Training Phase Algorithms}

The training algorithm for Efficient Destination Prediction can be broken down into two parts: the first phase initializing the total transition probabilities for all of the origin-destination pairs (Algorithm \ref{algInitialDPRTraining}), and the second one which continuously enhances our prediction by frequently updating the model (Algorithm \ref{algFrequentUpdateofDPR}). After the initialization of the total transition probabilities, we can proceed to continuously improve our model in a timely manner.

\begin{algorithm}  
\algsetup{linenosize=\small}
\scriptsize                  
	\caption{Initial Efficient Destination Prediction Training}         
	\label{algInitialDPRTraining}
	\textbf{Input:} the single-step Markov transition matrix $M^1$ \\
	\textbf{Output:} the total transition probabilities for all of the origin-destination pairs.
	\begin{algorithmic}[1]
	\algsetup{linenosize=\small}
	\scriptsize
		\FOR{$l=1$ to $2g$}
		\STATE{}\COMMENT{Compute every ETP whose distance is $\mathrm{l}$ for all the origin-destination pairs.}
		\STATE{
			Obtain $\mathrm{ETP}(i,j,l)$. (Equation 8)
		}
		\STATE{$p_{i\to j} = \mathrm{ETP}(i,j,l)$ }
		\ENDFOR
			\algsetup{linenosize=\small}
			\scriptsize
		\STATE{}\COMMENT{Obtain the TPD for each trip involving a detour the increment of length of which is always two to avoid unreachable cases whose transition probabilities are destined to be zero.}
			\algsetup{linenosize=\small}
			\scriptsize
		\STATE{$l=2$}
			\algsetup{linenosize=\small}
			\scriptsize
		\WHILE{$l<l_d$}
		\STATE{}\COMMENT{Compute every TPD whose length of detour is l for all the origin-destination pairs.}
		\STATE{
			$\mathrm{TPD}\left(i,j,l_{ij}+l\right)=\sum^4_{m=1}{\mathrm{TPD}(i,j_l,l_{ij_l})\mathrm{SSTP}(j_m,j)}$
		}
		\STATE{$p_{i\to j} += \mathrm{TPD}\left(i,j,l_{ij}+l\right)$}
		\STATE{$l+=2$}
		\ENDWHILE
	\end{algorithmic}
\end{algorithm}

The Initial EDP Training (Algorithm \ref{algInitialDPRTraining}) first obtains the ETP regarding two locations $i$ and $j$ whose corresponding L1 distance is $l$. Then we can yield the TPD pertaining to a detour of  $l$ in addition to the $l_{ij}$ distance associated with location $i$ and $j$. Note that when $l=0$ the TPD and ETP with respect to $i$ and $j$ are essentially identical. Hence ETP is first calculated in order to find TPD. We call this strategy 'Efficient First Detours Later' which enables us to obtain TPD in an efficient iterative dynamic-programming like manner.

When we compute the values of TPD, the increment of $l_{ij}$ distance is always two. This is because the value of one of the two TPDs whose difference of $l_{ij}$ is 1 must be 0 according to Theorem 1. Subsequently, we store the sum of TPDs in the corresponding total transition probability $p_{i\to j}$. 

\begin{algorithm}  
\algsetup{linenosize=\small}
\scriptsize                 
	\caption{Frequent Update of the Efficient Destination Prediction Model}         
	\label{algFrequentUpdateofDPR}
	\textbf{Input:} the set $G_c$ of cells that have undergone changes causing the alteration of their transition probabilities. \\
	\textbf{Output:} the updated total transition probabilities for all of the origin-destination pairs.
	\begin{algorithmic}[1]
	\algsetup{linenosize=\small}
	\scriptsize 
		\FOR{every cell $g_i$ in the map}
		\algsetup{linenosize=\small}
		\scriptsize
		\STATE{}\COMMENT{Identify the closest Obligatory Transit Point in relation to $g_i$}
		\algsetup{linenosize=\small}
		\scriptsize
		\STATE{
			Find the nearest OTP in $G_c$, denoted by $\mathrm{OTP}_{g_i}$.
		}
		\algsetup{linenosize=\small}
		\scriptsize
		\STATE{}\COMMENT{Identify the Transition Affected Area with regard to $\mathrm{OTP}_{g_i}$}
		
		\algsetup{linenosize=\small}
		\scriptsize\STATE{
			Find $\mathrm{TAA(}g_i,\mathrm{OTP}_{g_i}\mathrm{,D,T)}$, denoted by $\mathrm{TAA(OTP}_{g_i}\mathrm{)}$.
		}
		\algsetup{linenosize=\small}
		\scriptsize
		\FOR{every cell $g_j$ in $\mathrm{TAA(}\mathrm{OTP}_{g_i}\mathrm{)\ }$}
	\algsetup{linenosize=\small}
	\scriptsize
		\STATE{
			$\mathrm{ETP}(g_i,g_j,l)=\mathrm{ETP}(g_i,A^1_{ij},l-1)\times \mathrm{SSTP}(A^1_{ij},j)+\mathrm{ETP}(g_i,A^2_{ij},l-1)\times \mathrm{SSTP}(A^2_{ij},j)$
		}
		\STATE{$p_{i\to j}=\mathrm{ETP}(g_i,g_j,l)$}
		\algsetup{linenosize=\small}
		\scriptsize
		\STATE{$l=2$}
		\algsetup{linenosize=\small}
		\scriptsize
		\WHILE{$l<l_d$}
		\algsetup{linenosize=\small}
		\scriptsize
		\STATE{}\COMMENT{Compute every TPD whose length of detour is l for all the origin-destination pairs.}
		\algsetup{linenosize=\small}
		\scriptsize
		\STATE{
			$\mathrm{TPD}\left(i,j,l_{ij}+l\right)=\sum^4_{m=1}{\mathrm{TPD}(i,j_l,l_{ij_l})\mathrm{SSTP}(j_m,j)}$
		}
		\algsetup{linenosize=\small}
		\scriptsize
		\STATE{$p_{i\to j}+=\mathrm{TPD}\left(i,j,l_{ij}+l\right)$}
	\algsetup{linenosize=\small}
	\scriptsize
		\STATE{$l+=2$}
		\ENDWHILE
		\ENDFOR
		\ENDFOR
	\end{algorithmic}
\end{algorithm}

Similarly, when we update the EDP model (Algorithm \ref{algFrequentUpdateofDPR}) we first compute the values of ETPs residing within a TAA. And then we move on to find their corresponding TPDs and keep track of  the sum in $p_{i\to j}$. The step size of this loop is still 2 in accordance with Theorem 1.

\begin{algorithm}  
	\algsetup{linenosize=\small}
	\scriptsize                  
	\caption{Find $\mathrm{TAA(}g_i,\mathrm{OTP}_{g_i}\mathrm{,D,T)}$}         
	\label{algFindTAA}
	\textbf{Input:} cell $g_i$, $\mathrm{OTP}_{g_i}$, the set of detours $\mathrm{D}$.\\	
	\textbf{Output:} the corresponding $\mathrm{TAA(}g_i,\mathrm{OTP}_{g_i}\mathrm{,D,T)}$ regarding cell $g_i$, detours $\mathrm{D}$ and total travel distances $\mathrm{T}$.
	\begin{algorithmic}[1]
		\algsetup{linenosize=\small}
		\scriptsize
		\FOR{every $o\in\mathrm{OTP}_{g_i}$}
			\STATE{}\COMMENT{First obtain the initial rectangular transition affected area with respect to $i$ and $j$}
			\STATE{$\mathrm{TAA(}g_i,o\mathrm{,D,T)}\leftarrow rect(i,j)$}
			\FOR{every $d\in D$}
				\STATE{}\COMMENT{Then expand the previous transition affected area by taking in its border neighbors}
				\algsetup{linenosize=\small}
				\scriptsize
				\STATE{$\mathrm{TAA(}g_i,o,d\mathrm{,T)}\cup \mathrm{TAA(}g_i,o,d-2\mathrm{,T)}.borderNeigbors(i,j) $}
			\ENDFOR
		\ENDFOR
	\end{algorithmic}
\end{algorithm}

The TAA of interest can be found by first identifying the initial rectangular area with respect to cell $i$ and its OTP -- $j$ (line 3, Algorithm \ref{algFindTAA}). This rectangle essentially comprises all the cells whose detours are 0 and can be obtained as follows: first we draw a vertical line and then a horizontal one across $j$; the whole map now is partitioned into 4 regions; then the rectangle that is in the diagonal direction of the region containing cell $i$ is the desirable $rect(i,j)$. Once we have initialized this 0-detour TAA, we can then move on to find TAAs with longer detours by gradually extending their smaller counterparts through taking in their border neighbors as is shown in Figure \ref{Figure7TAAandTAAExtended} (line 6, Algorithm \ref{algFindTAA}).

\section{Experimental Evaluation}
We assess our algorithm and its competitors in this section. The dataset and the evaluation criteria are first described. Then we present their running time and the effectiveness of responding to the queries. Specifically, the study on run-time efficiency concerns both the time for model training and query answering. All the experiments are conducted on a desktop computer with 4GB of memory and a quad core 2.7GHz CPU.

\begin{figure}[tbph]
	\centering
	\includegraphics[scale=0.43]{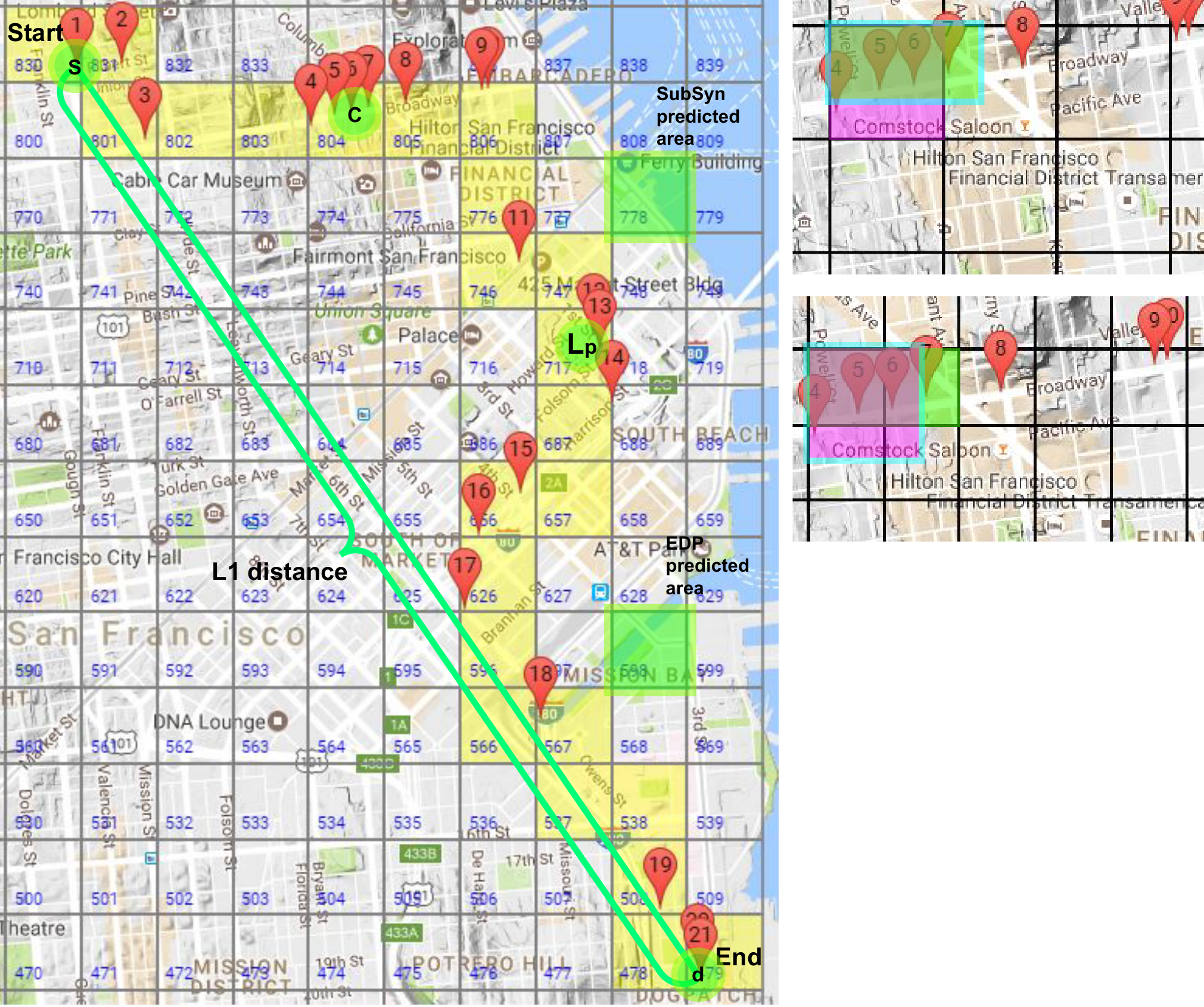}
	\caption{An example illustrating the effectiveness of our approach.}
	\label{RealExample}
\end{figure}

Shown in Figure \ref{RealExample} (on the left) is an example demonstrating the benefit of our approach which predicts an area that is closer to the final destination.

Starting from location 1, the trajectory ends at location 21. Assume that the driver is now at location 6. Our baseline predicts that the destination is somewhere around location 11 (cell 778). Differing from the SubSyn series algorithm which considers only two discrete locations ( location 1 and location 6 in this case), our approach draws on the route information concerning the two places and discovers the most probable future one to be location 13 ($\mathrm{L_p}$). And this inference leads us to the prediction that is closer to the true destination. Also notice that the route (length=24) taken by the driver is longer than the L1 distance (length=22) between the origin-destination endpoints. This confirms the effectiveness of our incorporation of detour distances.

The distinction between the two nearest cells gets increasingly blurred when the granularity grows. In the example shown in Figure \ref{RealExample} (on the right), it is apparent that the cell (light purple) residing in the 40x40 map (top) is roughly identical to the combined region of the two cells (light green) in the 60x60 map (bottom). The two regions mutually covers a large area of one another. This substantial degree of overlap between the two demonstrates why 2nd order Markov model (SubSynEA) tends to bring about limited improvement to the baseline SubSyn or SubSynE.

\textbf{Baseline Algorithms.} SubSynE and SubSynEA are used in our experiment to illustrate two facets of our algorithm -- the capacity to efficiently train the model that considers detour distances (EDP V.S. SubSynE) and the benefits accruing from both the constant updating of the model and the integration of detours (EDP V.S. SubSynEA). 

\textbf{Datasets.} We test all the algorithms on a real-world dataset that is openly available and on a synthetic one.

\textit{Real-World Data:} This dataset \cite{ge2010energy} encompasses the GPS location records pertaining to the real-time whereabouts of 514 taxis running in San Francisco Bay Area during a time span of 30 days. 10000 trajectories are randomly selected as the queries submitted by users while the remaining portion serves as the training set. This dataset is used to evaluate both the prediction accuracy and training efficiency of all the methods.

\textit{Synthetic Data:} We generate one synthetic data set in the form of single-step Markov matrix filled with transition probabilities. This dataset is solely for the assessment of training efficiency. Its size corresponds to the respective granularity of a real-world dataset.

\begin{figure}[t]
	\centering
	\includegraphics[scale=0.36]{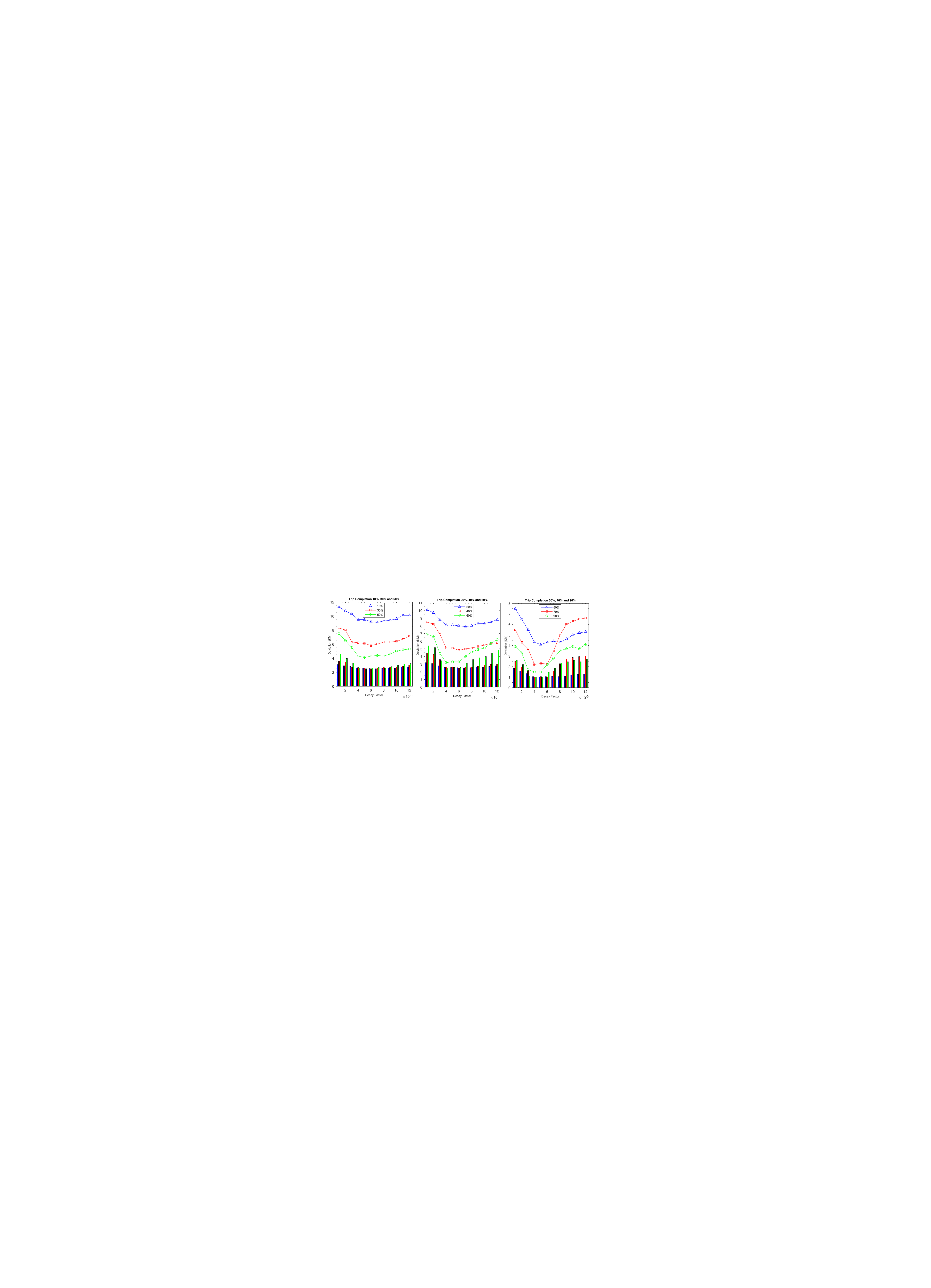}
	\caption{The effect of decay factor.}
	\label{DecayFactor}
\end{figure}

\textbf{The Effect of Decay Factor.} The decay factor determines the speed of decline of the predicted percentage. Figure \ref{DecayFactor} plots the deviation against the decay factor. The bars at the bottom of this chart, with three bars regarding their respective completion point in a group standing close, indicate the resulted difference between one decay factor and the one that yields the least deviation.

Our experiment shows that, at the earlier part of a trip, a larger decay factor is preferred. As the trip gradually draws to its end, a smaller $\alpha$ makes a more favorable choice. We set this parameter to 0.004 to strike the right balance.

\subsection{Efficiency of Training Algorithm}
\begin{figure}
	\includegraphics[scale=0.55]{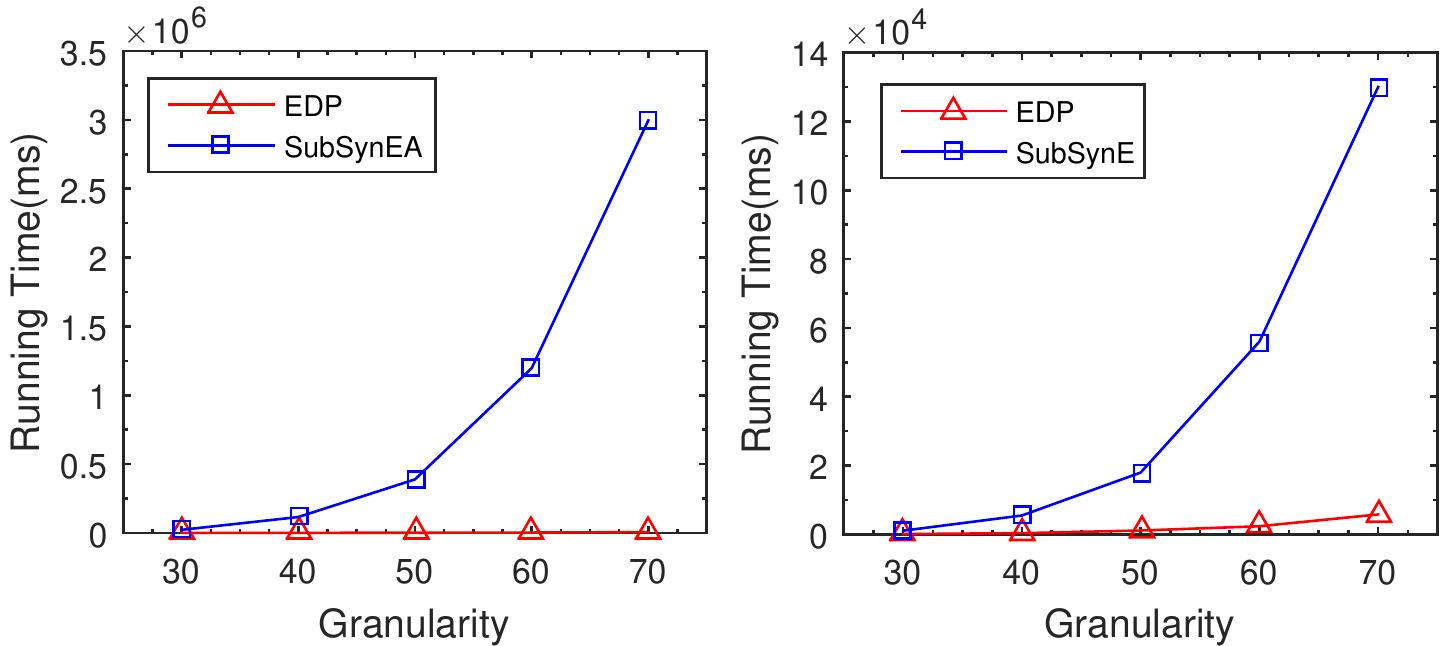}
	\caption{Training time of EDP, SubSynE and SubSynEA.}
	\label{Figure13RunTimeHVSSE}
\end{figure}

As is shown in Figure \ref{Figure13RunTimeHVSSE}, it is apparent that our approach is more superior to SubSynE and SubSynEA in terms of training time as the granularity rises. This is particularly true once the map becomes more fine-grained. Our approach is 16.1 times and 348.5 times faster than SubSynE and SubSynEA respectively when we set the granularity to 50. The training time of both SubSynE and SubSynEA has soared even more dramatically after the granularity exceeds 50. The zero entries most of the time constitute far beyond 50\% of the elements of transition matrix for SubSynE and SubSynEA. They repeatedly go through the process of data retrieval from main memory (cache will simply not fit owing to the enormous amount of them), double-precision floating point multiplication and storing them back. This process imposes an onerous yet unnecessary burden on the overall efficiency. A\label{key}fter the preceding computation is done, it often just yields another zero that contributes little to the computation of non-zero transition probabilities but actively involves in yet another vicious cycle of this sort. Furthermore, the inadvertent inclusion of detour distances dictated by matrix multiplication significantly exacerbates the performance of SubSynEA (Figure \ref{Figure13RunTimeHVSSE}).

\subsection{Evaluation of Prediction Algorithm}

\subsubsection{Efficiency Evaluation} 
In the phase of responding to user queries, our solution needs the incorporation of semi-lazy trajectory preprocessing to locate the most likely future position $\mathrm{L_p}$. Thus it takes extra time for our approach to factor into this determinant. This trade-off is favorable since it additionally incurs merely a fraction of a second (around an extra 65ms in most cases) but vastly improves the accuracy. The succeeding step of destination prediction can be performed very fast (around 0.05ms) as it simply retrieves the numeric values of transition probabilities for further calculation and comparison. The predominant factor of these algorithms is the training time which sets our algorithm apart from its competitors.

\subsubsection{Accuracy Evaluation} 
First we would like to discuss about the measures we use to gauge the performance. Two specific locations in the course of a trip -- the 30\% and 70\% completion point - particularly draw our attention as they indicate how well an algorithm will fare soon after a traveler just begins his trip or soon before he arrives at his destination. Moreover, the impact of grid granularity is of our concern since it correlates strongly with the effectiveness of our approach. Besides we also alter the completion percentage of trip and the ratio of identical trajectories (shown in Figure \ref{MatchRatio}). Here identical trajectories refer to those in the testing dataset that are perceived as the same with their counterparts in the training dataset. Judged against the yardstick of this ratio, all the algorithms can be examined from a more practical perspective which reflects their capability of dealing with the recurring historical data (i.e. exact matches) as well as generalizing to completely novel scenarios (i.e. new routes that emerge for the first time).

\begin{figure}[t]
	\centering
	\includegraphics[scale=0.53]{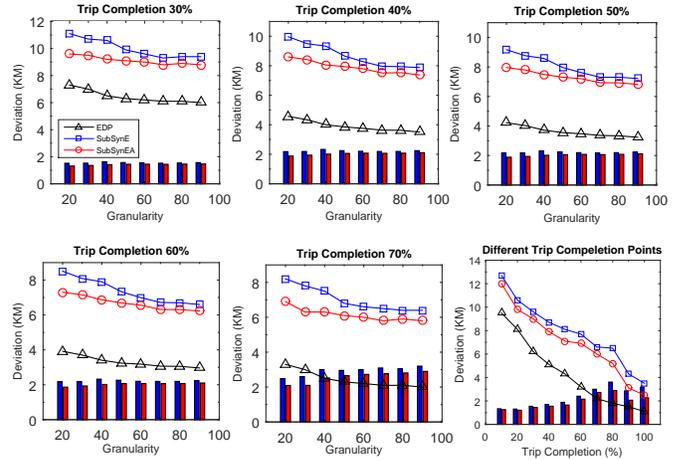}
	\caption{Average deviation from destination at the 30\% - 70\% completion point and at the different trip completion points.}
	\label{Figure15Deviation30}
\end{figure}

\textit{30\% - 70\% Completion Points: }
We compute the mean of the deviation distances of the top three destinations given by the algorithms. Our solution consistently outperforms SubSyn and SubSynEA in terms of prediction accuracy quantified by the average deviation from the ground truth (Figure \ref{Figure15Deviation30}). The granularity of a map plays an essential role in improving the accuracy though this effect gradually dwindles as the map becomes more fine-grained. Moreover, SubSynEA (second-order model) produces much better prediction results than SynSynE (first-order model) in the more coarse-grained settings. This distinction slowly fades away as the granularity increases, which is in agreement with our earlier analysis that the second-order model is prone to degradation since the rising amount of cells in a map obscures the distinction of two geographically isolated locations (Figure \ref{RealExample} on the right). 

\textit{Different Completion Points: } The ability to pinpoint the cause of transitioning variation (OTP and TAA) and to address this problem by only re-computing the affected transition probabilities makes our approach very efficient and more accurate. The synergy of the two aforementioned mechanisms gives rise to the definite edge of our approach over its competitors in terms of prediction accuracy, which is particularly evident during the course starting from the 25\% completion point and ending at the 85\% completion point, the primary stage for location prediction (Figure \ref{Figure15Deviation30}). The potential opportunities for various tasks such as POI (point of interest) recommendation and advertising abound especially in this course..

\begin{figure}[t]
	\centering
	\includegraphics[scale=0.45]{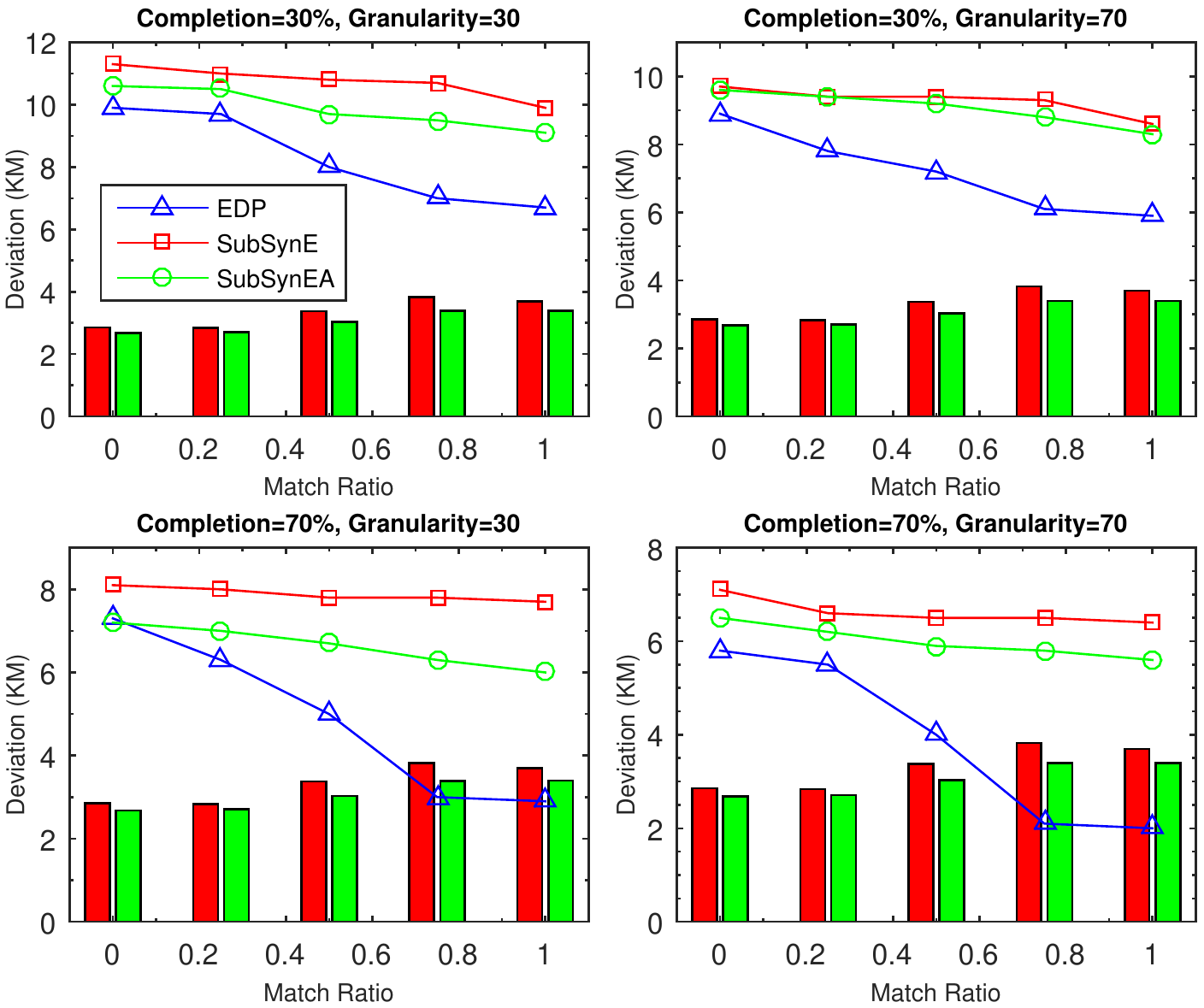}
	\caption{Prediction accuracy varies with different match ratios.}
	\label{MatchRatio}
\end{figure}

\subsection{Analysis of Accuracy Improvement}

The second-order Markov model underlying SubSynEA enhances the prediction accuracy at the expense of a substantial increase in transition states which are offset to some extent by sparse matrix multiplication. Moreover, SubSynEA excludes the consideration for detour distances favoring the simplification of the model. Rather than employing a high-order Markov model, we stick with the first-order model and apply the semi-lazy path prediction algorithm first to discover the most probable future location. We find that our solution outperforms its competitors since the route picked by a user should be best described by the model focusing on the trajectory itself as opposed to several discrete Markov states. 

The lack of a proper way handling the user-chosen route undermines the chances of right prediction of SubSyn. SubSynEA attempts to remedy this problem by only additionally considering the nearest historical location. However, a similar issue will arise regarding this technique that the itinerary traveled so far is still partially represented by merely three locations. The effectiveness of this strategy gradually diminishes as the granularity of grid map becomes less coarse. The distinction of the two states associated with the neighboring region of current location gets increasingly blurred, implicating that SubSynEA has the inherent propensity to fall back into SubSyn in fine-grained settings. This may well account for its mediocre performance under such conditions in our experiment.

\section{Conclusion}
In this paper we propose an efficient scheme for destination prediction that runs an order of magnitude faster and gains an increase of over 30\% in accuracy, compared with the state-of-the-art approach. Our solution mainly involves the inclusion of semi-lazy prediction, the optimization of the Markov transition matrix multiplication and a feasible frequent update method for our model. Experimental results with respect to the preceding two dimensions of our work have demonstrated its efficiency and effectiveness. 

\textbf{Acknowledgments.} The authors would like to thank Prof. Xifeng Yan for his valuable comments.



\appendix
This section is devoted to the mathematically rigid analysis of the reduction in time and space complexity of our approach.

\begin{figure}
	\centering
	\subfigure[]{\includegraphics[width=0.20\textwidth]{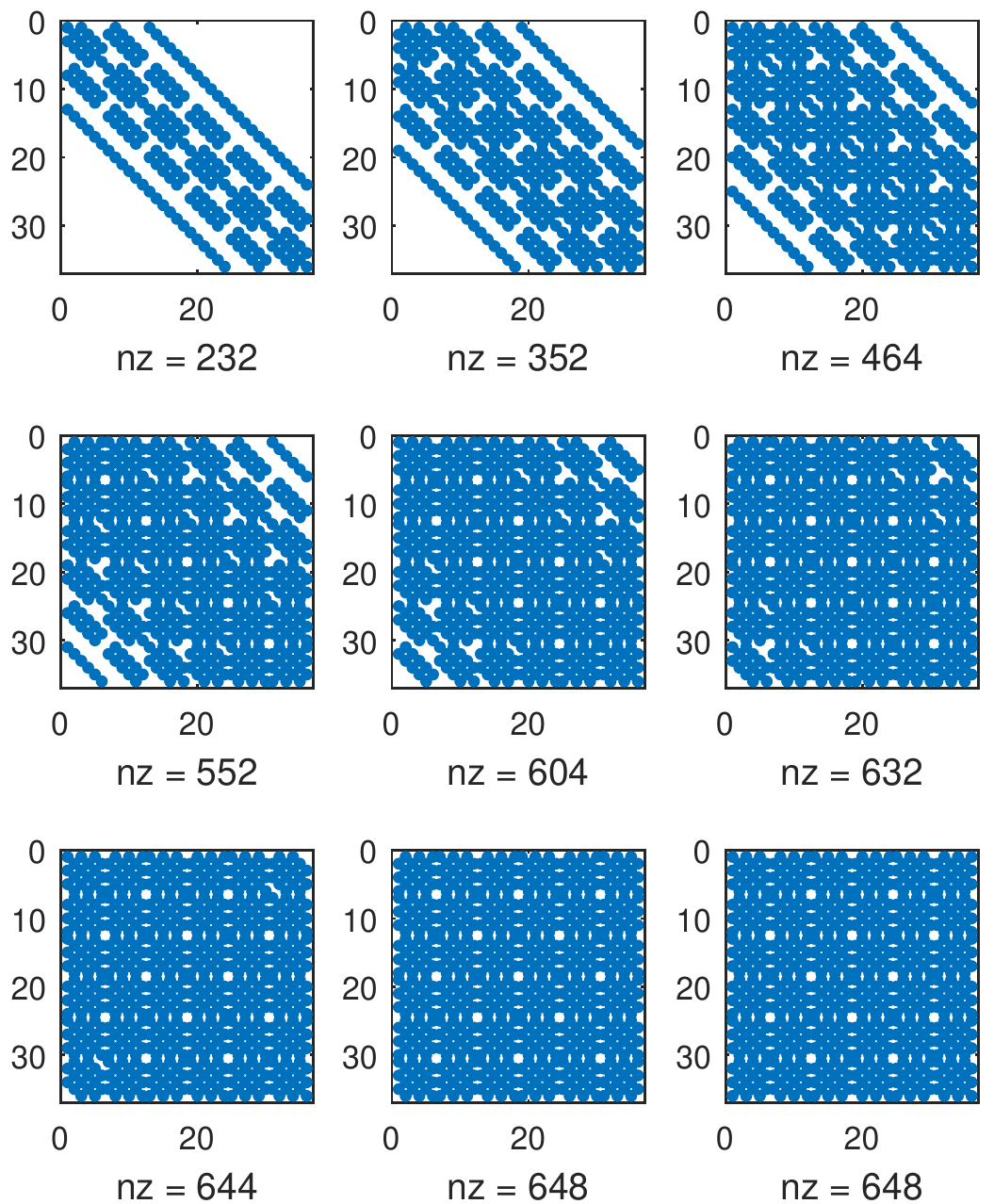} \label{NonZeroSMM}}
	\hspace{0.6cm}
	\subfigure[]{\includegraphics[width=0.204\textwidth]{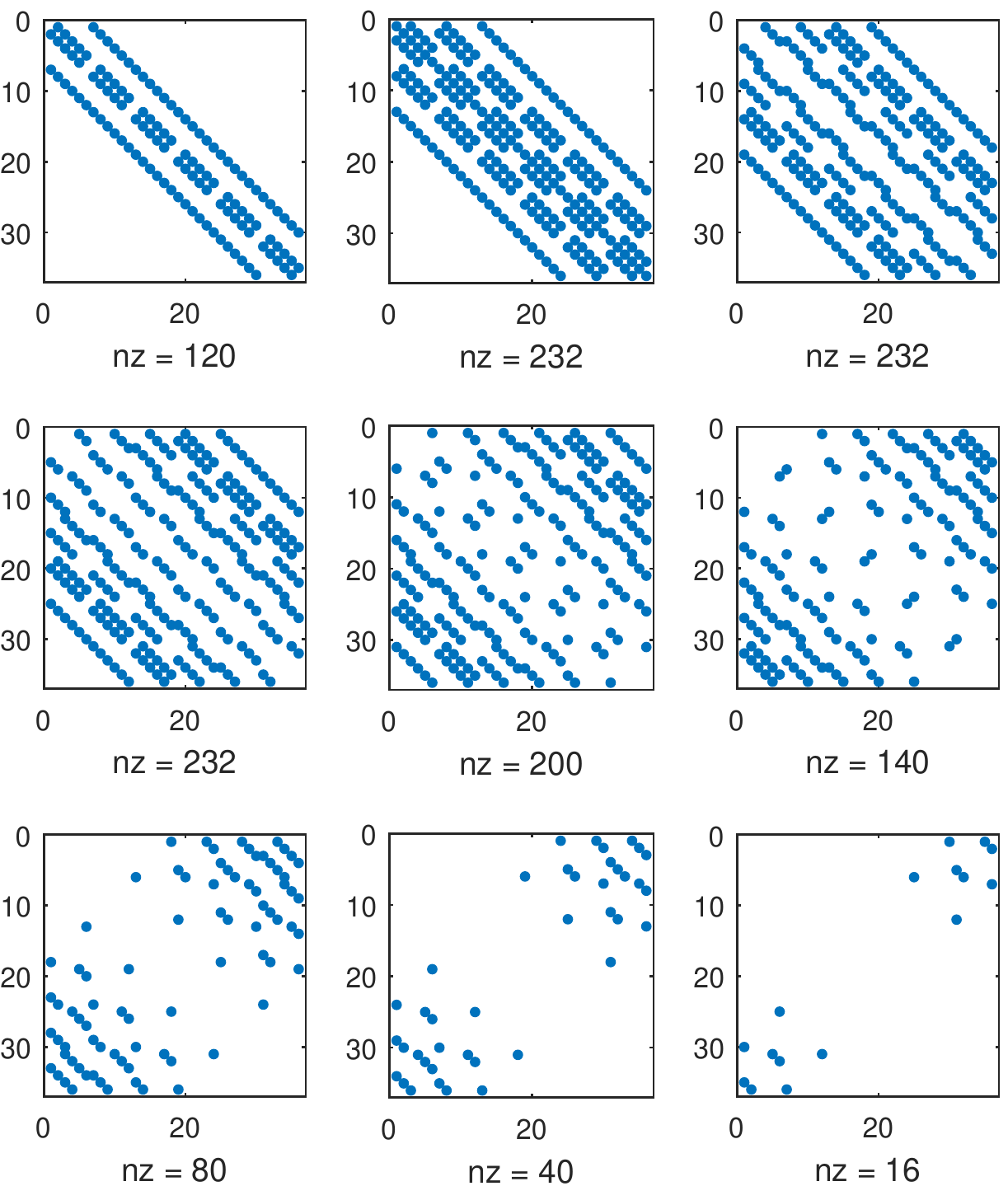} \label{NonZeroETP}}
	\caption{Non-zero elements of SMM and ETP}
\end{figure}

The transition matrix of the prediction model can be shown in Figure \ref{NonZeroSMM} and Figure \ref{NonZeroETP} for Sparse Matrix Multiplication and our approach respectively. It is apparent that the matrix gradually becomes denser with the increase of the transition steps for SMM. To make it more amenable for further analysis, we partition the whole matrix into $n$ blocks. The evolving pattern of each submatrix then can be analyzed: starting from the major diagonal, the non-zero elements gradually spread through the entire block, and in the end occupy half of the locations in an alternating manner.
To yield the sum of non-zero elements of the whole transition matrix, we first obtain this sum of each block. Let $i$ be the step of transition and $m$ the distance of the diagonal deviating from the major one. Then the amount of non-zero elements of a diagonal of a submatrix can be given by

{\small
\[{\theta }_{i,m}=\left[\sum^{i-m}_{j=0}{{\lambda }_{i+j+m}\mathrm{(}\sqrt{n}-j)}\right]-\frac{{\lambda }_{i+m}\sqrt{n}}{2},i<\sqrt{n},\]
}where ${\lambda }_a\mathrm{=}1+{(-1)}^a$ denotes the alternating factor that decides whether the entries of a diagonal is zero.

Once the non-zero elements has spread throughout the whole block (i.e. after $\sqrt{n}$ steps of transition have been taken), ${\theta }_{i,m}$ fluctuates between two numbers both of which are around half of the matrix size

{\small
\[{\theta }_{i,m}=\frac{1}{2}({\lambda }_{i-1}{\theta }_{\sqrt{n},m}+{\lambda }_i{\theta }_{\sqrt{n}-1,m})\mathrm{,\ }i\ge \sqrt{n}.\] 
}

Given the current step of transition $i$, we can sum up the number of non-zero entries of each submatrix by

{\small
\[Z_{SMM}\mathrm{(}i\mathrm{)=}{\sqrt{n}\theta }_{i,0}+2\sum^{\mathrm{t}}_{m=1}{\left(\sqrt{n}-m\right){\theta }_{i,m}},\] 
}

{\small\[t=\left\{ \begin{array}{c}
i\ \ \ \ \ \ \ \ \ \ \ (i<\sqrt{n}) \\ 
\sqrt{n}-1\ \ \ (i\ge \sqrt{n}) \end{array}
\right.\] 
}where the variable upper bound $t$ imposes the limit of the number of summation of a block once the transition steps exceed its boundary.
%

Due to the involvement of ${\lambda }_a$, we have to discuss each case with respect to the parity $\alpha$.

First we expand the term ${\theta }_{i,m}$:

\textbf{Case 1}: $i<\sqrt{n}\mathrm{,\ }$ and $i$ and $m$ have the same parity, which results in an even-number sum $i+m$ and an even-number difference $i-m$.
{\small
\[{\theta }_{i,m}=\left[\sum^{i-m}_{j=0}{{\lambda }_{i+j+m}\left(\sqrt{n}-j\right)}\right]-\frac{{\lambda }_{i+m}\sqrt{n}}{2}\] 
}

{\small\[=\left(\sum^{\frac{i-m}{2}}_{j=0}{\sqrt{n}-2j}\right)-\frac{\sqrt{n}}{2}\] }

{\small\[\mathrm{=}\left(\frac{i-m}{2}+1\right)\left(\frac{m-i}{2}\mathrm{+}\sqrt{n}\right)\mathrm{-}\frac{\sqrt{n}}{2}\]
}

\textbf{Case 2}: $i<\sqrt{n}\mathrm{,\ }$ and $i$ and $m$ have different parities, which results in an odd-number sum $i+m$ and an odd-number difference $i-m$.
{\small
\[{\theta }_{i,m}=\left[\sum^{i-m}_{j=0}{{\lambda }_{i+j+m}\left(\sqrt{n}-j\right)}\right]-\frac{{\lambda }_{i+m}\sqrt{n}}{2}\] 
}

{\small\[=\left(\sum^{\frac{i-m+1}{2}}_{j=1}{\sqrt{n}-2j}+1\right)-\frac{\sqrt{n}}{2}\] 
}

{\small\[\mathrm{=}\frac{1}{4}\left(i-m+1\right)\left(2\sqrt{n}-i+m-1\right)-\frac{\sqrt{n}}{2}\]
}
\textbf{Case 3}: $i\ge \sqrt{n},$ we simply substitute $i$ with $\sqrt{n}$ or $\sqrt{n}-1$, the parity of $i$ and $\sqrt{n}$ jointly determine the respective case.

As for ${\theta }_{i,0}$ we have

{\small\[{\theta }_{i,0}=\left\{ \begin{array}{c}
\left(\frac{i}{2}+1\right)\left(\sqrt{n}-\frac{i}{2}\right)-\frac{\sqrt{n}}{2}\ \ \ \ \ \ \ i\ is\ an\ even\ number \\ 
\frac{1}{4}\left(i+1\right)\left(2\sqrt{n}-i-1\right)-\frac{\sqrt{n}}{2}\ \ \ \ \ \ \ \ \ \ \ \ \ \ \  otherwise \end{array}
\right.\]
}

Then the summation involving the variable upper bound $t$ can be expanded by:

{\small\[\sum^{\mathrm{t}}_{m=1}{\left(\sqrt{n}-m\right){\theta }_{i,m}}\] 
}

\textbf{Case 1}: $t$\textit{ }is an odd number $t=i<\sqrt{n}$

{\small\[\mathrm{=}\sum^{\frac{\mathrm{t+1}}{\mathrm{2}}}_{m=1}{\left(\sqrt{n}-2m+1\right){\theta }_{i,2m-1}}+\sum^{\frac{\mathrm{t-1}}{\mathrm{2}}}_{m=1}{\left(\sqrt{n}-2m\right){\theta }_{i,2m}}\] 
}

\textbf{Case 2}: $t$\textit{ }is an even number $t=i<\sqrt{n}$

{\small\[\mathrm{=}\sum^{\frac{\mathrm{t}}{\mathrm{2}}}_{m=1}{\left(\sqrt{n}-2m+1\right){\theta }_{i,2m-1}}+\sum^{\frac{\mathrm{t}}{\mathrm{2}}}_{m=1}{\left(\sqrt{n}-2m\right){\theta }_{i,2m}}\]
}

\textbf{Case 3}: When $t=\sqrt{n}-1\mathrm{\ }(i\ge \sqrt{n})\mathrm{\ }$we simply substitute $t$ with $\sqrt{n}-1$ in the preceding two equations, and the parity of $\sqrt{n}-1$ determines the respective case.

From the above analysis it is evident that ${\theta }_{i,m}=O(i^2)$, and $Z_{SMM}\left(i\right)=O(i^4)$ can be derived from the expansion of the term $\sum^{\mathrm{t}}_{m=1}{\left(\sqrt{n}-m\right){\theta }_{i,m}}$ by summing the sequence of cubes regarding $i$.

Next we study the non-zero elements dealt with by our approach. Likewise, first we denote the number of non-zero entries in a block by

{\small\[{\delta }_m=\left\{ \begin{array}{c}
\sqrt{n}\ \ \ \ \ \ \ \ \ \ \ \ \ m=0 \\ 
2\left(\sqrt{n}-m\right)\ \ \ \ \ m<\sqrt{n} \\ 
0\ \ \ \ \ \ \ \ \ \ \ \ \ \ \ \ \ \ \ m\ge \sqrt{n} \end{array}
\right.\]
}

And the total amount of elements of ETP(Efficient Transition Probability) can then be given by:

{\small\[Z_{ETP}\left(i\right)=\sqrt{n}{\delta }_i+2\sum^i_{j=1}{\left(\sqrt{n}-j\right){\delta }_{i-j}},i\mathrm{\in }\left[\mathrm{1,\ 2}\sqrt{n}\right]\] }

It is worth mentioning that ${\theta }_{i,m}$ for $Z_{SMM}\mathrm{(}i\mathrm{)}$ gradually grows with $i$ and peaks once $i$ exceeds $\sqrt{n}$. Unlike ${\theta }_{i,m}$ for $Z_{SMM}\mathrm{(}i\mathrm{)}$, ${\delta }_m$ for $Z_{ETP}(i)$ first increases with $i$ and then declines. We expand the term $\sum^i_{j=1}{\left(\sqrt{n}-j\right){\delta }_{i-j}}$ and examine its upper bound.

\textbf{Case 1}: $i<\sqrt{n}$
{\small\[\sum^i_{j=1}{\left(\sqrt{n}-j\right){\delta }_{i-j}}\] 

\[\mathrm{=2}\left(\sqrt{n}-i\right)\sqrt{n}+\sum^{i-1}_{j=1}{2\left(\sqrt{n}-j\right)\left(\sqrt{n}-i+j\right)}\]
}

\textbf{Case 2}: $i\ge \sqrt{n}$
{\small\[\sum^i_{j=1}{\left(\sqrt{n}-j\right){\delta }_{i-j}}\] 

\[\mathrm{=}\sum^{\sqrt{n}}_{j=i-\sqrt{n}+1}{2\left(\sqrt{n}-j\right)\left(\sqrt{n}-i+j\right)}\] 
}

The preceding analysis shows that $\sum^i_{j=1}{\left(\sqrt{n}-j\right){\delta }_{i-j}}=O(i^3)$ for both of the two cases through the summation of the series concerning $i$. Hence it is apparent that $Z_{ETP}\left(i\right)=O(i^3)$. Therefore, compared with sparse matrix multiplication, the reduction in time and space complexity of our approach is pronounced.


\begin{thebibliography}{10}
	
	\bibitem{do2012contextual}
	T.~M.~T. Do and D.~Gatica-Perez.
	\newblock Contextual conditional models for smartphone-based human mobility
	prediction.
	\newblock In {\em UbiComp}, pages 163--172, 2012.
	
	\bibitem{evensen2011mobile}
	K.~Evensen, A.~Petlund, H.~Riiser, P.~Vigmostad, D.~Kaspar, C.~Griwodz, and
	P.~Halvorsen.
	\newblock Mobile video streaming using location-based network prediction and
	transparent handover.
	\newblock In {\em Proceedings of the 21st international workshop on Network and
		operating systems support for digital audio and video}, pages 21--26, 2011.
	
	\bibitem{gao2014elastic}
	X.~Gao, B.~Firner, S.~Sugrim, V.~Kaiser-Pendergrast, Y.~Yang, and J.~Lindqvist.
	\newblock Elastic pathing: Your speed is enough to track you.
	\newblock In {\em UbiComp}, pages 975--986, 2014.
	
	\bibitem{ge2010energy}
	Y.~Ge, H.~Xiong, A.~Tuzhilin, K.~Xiao, M.~Gruteser, and M.~Pazzani.
	\newblock An energy-efficient mobile recommender system.
	\newblock In {\em KDD}, pages 899--908, 2010.
	
	\bibitem{gonzalez2007adaptive}
	H.~Gonzalez, J.~Han, X.~Li, M.~Myslinska, and J.~P. Sondag.
	\newblock Adaptive fastest path computation on a road network: a traffic mining
	approach.
	\newblock In {\em PVLDB}, pages 794--805, 2007.
	
	\bibitem{horvitz2012some}
	E.~Horvitz and J.~Krumm.
	\newblock Some help on the way: Opportunistic routing under uncertainty.
	\newblock In {\em UbiComp}, pages 371--380, 2012.
	
	\bibitem{jeung2008hybrid}
	H.~Jeung, Q.~Liu, H.~T. Shen, and X.~Zhou.
	\newblock A hybrid prediction model for moving objects.
	\newblock In {\em ICDE}, pages 70--79, 2008.
	
	\bibitem{krumm2006real}
	J.~Krumm.
	\newblock Real time destination prediction based on efficient routes.
	\newblock In {\em Society of Automotive Engineers (SAE) 2006 World Congress},
	volume~7, 2006.
	
	\bibitem{krumm2006predestination}
	J.~Krumm and E.~Horvitz.
	\newblock Predestination: Inferring destinations from partial trajectories.
	\newblock In {\em UbiComp}, pages 243--260. 2006.
	
	\bibitem{luo2013finding}
	W.~Luo, H.~Tan, L.~Chen, and L.~M. Ni.
	\newblock Finding time period-based most frequent path in big trajectory data.
	\newblock In {\em SIGMOD}, pages 713--724, 2013.
	
	\bibitem{mathew2012predicting}
	W.~Mathew, R.~Raposo, and B.~Martins.
	\newblock Predicting future locations with hidden markov models.
	\newblock In {\em UbiComp}, pages 911--918, 2012.
	
	\bibitem{monreale2009wherenext}
	A.~Monreale, F.~Pinelli, R.~Trasarti, and F.~Giannotti.
	\newblock Wherenext: a location predictor on trajectory pattern mining.
	\newblock In {\em SIGKDD}, pages 637--646, 2009.
	
	\bibitem{patterson2004opportunity}
	D.~J. Patterson, L.~Liao, K.~Gajos, M.~Collier, N.~Livic, K.~Olson, S.~Wang,
	D.~Fox, and H.~Kautz.
	\newblock Opportunity knocks: A system to provide cognitive assistance with
	transportation services.
	\newblock In {\em UbiComp}, pages 433--450. 2004.
	
	\bibitem{sametKDD16aircraft}
	H.~S. Samet~Ayhan.
	\newblock Aircraft trajectory prediction made easy with predictive analytics.
	\newblock In {\em KDD}, 2016.
	
	\bibitem{su2014crowdplanner}
	H.~Su, K.~Zheng, J.~Huang, H.~Jeung, L.~Chen, and X.~Zhou.
	\newblock Crowdplanner: A crowd-based route recommendation system.
	\newblock In {\em ICDE}, pages 1144--1155, 2014.
	
	\bibitem{wei2012constructing}
	L.-Y. Wei, Y.~Zheng, and W.-C. Peng.
	\newblock Constructing popular routes from uncertain trajectories.
	\newblock In {\em SIGKDD}, pages 195--203, 2012.
	
	\bibitem{xue2015solving}
	A.~Y. Xue, J.~Qi, X.~Xie, R.~Zhang, J.~Huang, and Y.~Li.
	\newblock Solving the data sparsity problem in destination prediction.
	\newblock {\em The VLDB Journal}, 24(2):219--243, 2015.
	
	\bibitem{xue2013destination}
	A.~Y. Xue, R.~Zhang, Y.~Zheng, X.~Xie, J.~Huang, and Z.~Xu.
	\newblock Destination prediction by sub-trajectory synthesis and privacy
	protection against such prediction.
	\newblock In {\em ICDE}, pages 254--265, 2013.
	
	\bibitem{xue2013desteller}
	A.~Y. Xue, R.~Zhang, Y.~Zheng, X.~Xie, J.~Yu, and Y.~Tang.
	\newblock Desteller: A system for destination prediction based on trajectories
	with privacy protection.
	\newblock {\em PVLDB}, 6(12):1198--1201, 2013.
	
	\bibitem{yuan2012discovering}
	J.~Yuan, Y.~Zheng, and X.~Xie.
	\newblock Discovering regions of different functions in a city using human
	mobility and pois.
	\newblock In {\em SIGKDD}, pages 186--194, 2012.
	
	\bibitem{chen2011discovering}
	C.~Zaiben, S.~H. Tao, and Z.~Xiaofang.
	\newblock Discovering popular routes from trajectories.
	\newblock In {\em ICDE}, pages 900--911, 2011.
	
	\bibitem{zheng2014modeling}
	J.~Zheng and L.~M. Ni.
	\newblock Modeling heterogeneous routing decisions in trajectories for driving
	experience learning.
	\newblock In {\em Proceedings of the 2014 ACM International Joint Conference on
		Pervasive and Ubiquitous Computing}, pages 951--961, 2014.
	
	\bibitem{zhou2013semi}
	J.~Zhou, A.~K. Tung, W.~Wu, and W.~S. Ng.
	\newblock A “semi-lazy” approach to probabilistic path prediction in dynamic
	environments.
	\newblock In {\em SIGKDD}, pages 748--756, 2013.
	
	\bibitem{ziebart2008navigate}
	B.~D. Ziebart, A.~L. Maas, A.~K. Dey, and J.~A. Bagnell.
	\newblock Navigate like a cabbie: Probabilistic reasoning from observed
	context-aware behavior.
	\newblock In {\em UbiComp}, pages 322--331, 2008.
	
\end{thebibliography}
\end{document}